\documentclass[preprint,showpacs,prb]{revtex4-1}
\usepackage{graphicx}
\usepackage{subfig}
\usepackage{dcolumn}
\usepackage{bm}
\begin{document}

\title{Ligand effects on the electronic structure and magnetism of magnetite surfaces}
\author{K.  Brymora} 
\author{F. Calvayrac }%
\email{ Florent.Calvayrac@univ-lemans.fr}
\affiliation{
 Institut des Mol\'ecules et Mat\'eriaux du Mans, PEC-IMMM, CNRS UMR 6283, PRES LUNAM
Universit\'e du Maine, Av.Messiaen, 72085 Le Mans Cedex 9, France
}

\date{\today}

\begin{abstract}

We address the effect of functionalization on the electronic
and magnetic properties of magnetite surface as an indicator of the
same properties in nanoparticles too big for a direct ab-initio approach.
Using well-established methods and references (namely LDA+U on
magnetite surfaces) we could verify the validity of our approach,
and using two typical ligands, dopamine and citrate, namely 
 $\pi$ and $\sigma$ electron donors,
we could predict that those ligands would induce a different
change in the electronic properties of the systems, but in both cases 
an enhancement of magnetization. 
\end{abstract}


\maketitle

\section{Introduction} 

During the last decade, strong efforts have been
devoted to  studying magnetic iron oxide nanoparticles, 
which present promising applications in various fields,
especially in medicine, for instance as 
contrast agents in MRI, drug delivery vectors, or as heat
mediators in hyperthermia treatments.
Maghemite ($\gamma$ Fe$_2$O$_3$) and magnetite
(Fe$_3$O$_4$ ) nanoparticles,
besides their low cost, 
high chemical stability  and low toxicity. 
 have extremely interesting properties
due to the high magnetic moments caused by  their
ferrimagnetism. Under a certain size,
those nanoparticles present  zero coercivity~\cite{Figuerola}  
which makes them particularly useful due to the apparition
of superparamagnetism and the prevention of the clogging
of particles. In hyperthermia treatments, the application
of an alternating field on top of a strong static magnetic field leads
to heat dissipation by hysteresis losses. Once again, a rather small size
of nanoparticles seems to be optimal, about 20 nm in the
case of magnetite, for instance~\cite{levy,levy2}. Some authors~\cite{Tang} have
even suggested the possibility of engineerinf nanoparticles to exhibit a
Curie temperature close to the therapeutic one so that their
heating power switches off above this desired temperature.

Besides, the question of the biocompatibility of  nanoparticles arise.
Various means of surface manipulation of the nanoparticles are used,
such as the use of ligands, layers of polymers or of other materials such as
proteins~\cite{Klem}.
The first goals  are to make the particles hydrophilic, then to make
them biocompatible, by reducing their sensibility to the pH of the solution
in which they are used. The next issue is to target them to tumors by
adding specific ligands~\cite{Thorek}. The question then arises
of the change induced on the physical properties of the nanoparticles
by the presence of a specific ligand.

Among those changes, the question of the charge order at the surface
of the nanoparticles  is a crucial
one since it can influence magnetic as well as conduction properties.
A series of recent papers has successfully addressed, for instance,
the theoretical and experimental descriptions of charge order and Verwey transition in bulk
magnetite~\cite{Piekarz},~\cite{Duffy} or nanocrystals~\cite{Lei} as well as as magnetite
surface~\cite{cheng},~\cite{Jordan},\cite{Lodziana}. 
Magnetite is at room temperature a poor metal, the electronic
conductivity being caused by the $t_{2g}$ orbitals contributed
by the so-called octahedral Fe(B), randomly distributed, cations. 
Various models have been proposed to describe the magnetite surface
~\cite{Pentcheva}, but recent DFT approaches~\cite{Spiridis,Lodziana,Mulakulari,Mulakulari2,Parkinson} seem
to be the most promising for iron oxide surfaces, as for iron oxide clusters~\cite{Lopez}.

In this paper, we  propose to extend those studies in order to address the role
of commonly used ligands on the charge order at the surface of magnetite and
their effect on the magnetic and electronic properties of this material.
If the radius of the nanoparticles used in medicine is large enough, we argue
that the local effects are indistinguishable from surface effects.

\section{Method and chosen systems}
We chose to describe the systems by density functional theory, using a well-established combination
of plane waves and pseudopotentials, since core electrons do not contribute to the 
phenomena we are interested in, and since periodicity is assumed in two directions at least
to describe surfaces, a large enough vacuum being added in the third direction in order to
minimize interactions in between periodic replicas. 

The magnetite surface was thus built by taking the unit cell from~\cite{ocd,refocd} 
adding 3.4 nm of vacuum in the [001] direction and structurally annealing it with the PWscf program from
the Quantum Espresso suite~\cite{QE-2009}. 

We used the  LDA+U method for magnetite (in order to have an state presenting a very small gap, as 
reported elsewhere in the literature~\cite{pinto}). We 
used  Marzari-Vanderbilt smearing and a  Gaussian smearing factor of 0.02.  
a 0.17 mixing factor for self-consistency was used. The LDA+U parameters
were set at U=4.5 eV for Fe and J=0 in accordance with previous papers~\cite{Lodziana,pinto}. We used
an automatic sampling of the first Brillouin zone. No symmetry was used. 
Due to the corresponding very high computational cost, at first we used an energy cut-off
of 27 Ry and turned off the spin degree of freedom in order to explore the structural
stability of the system, and progressively increased this cutoff to a value of 30 Ry
which we found not to perturb the results, as well as using a grid of 4X4X4 K-points in 
the first Brillouin zone. Correspondingly, we turned on the spin degree of freedom
in the calculation after having established stable structures, and even checked 
that the result was the same up to two decimals in various observables when using
non-collinear magnetism, showing that the systems are essentially collinear ferrimagnets.

\begin{figure}
{\includegraphics[scale=0.14,angle=0]{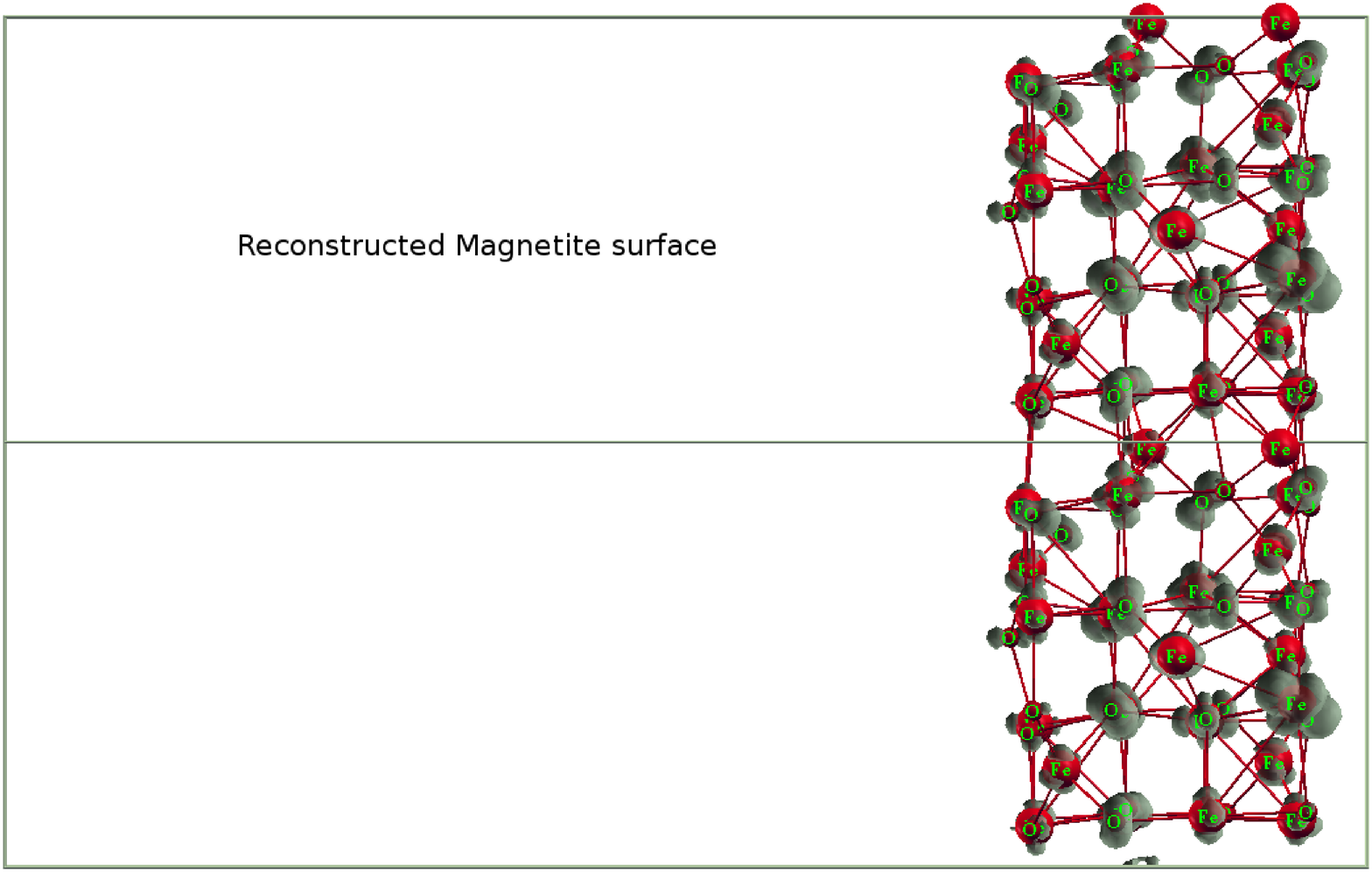}}
{\includegraphics[scale=0.11,angle=0]{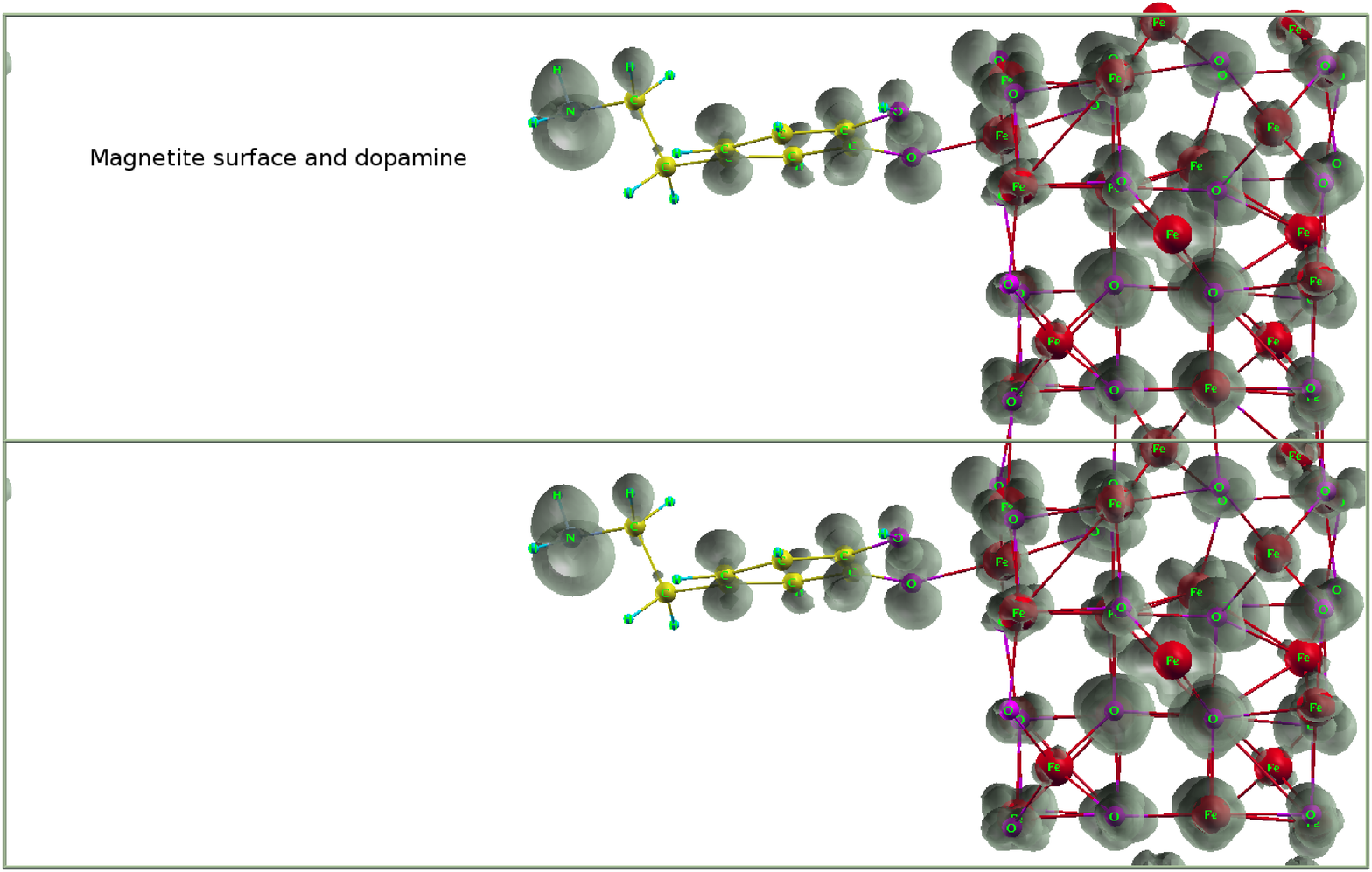}}
{\includegraphics[scale=0.12,angle=0]{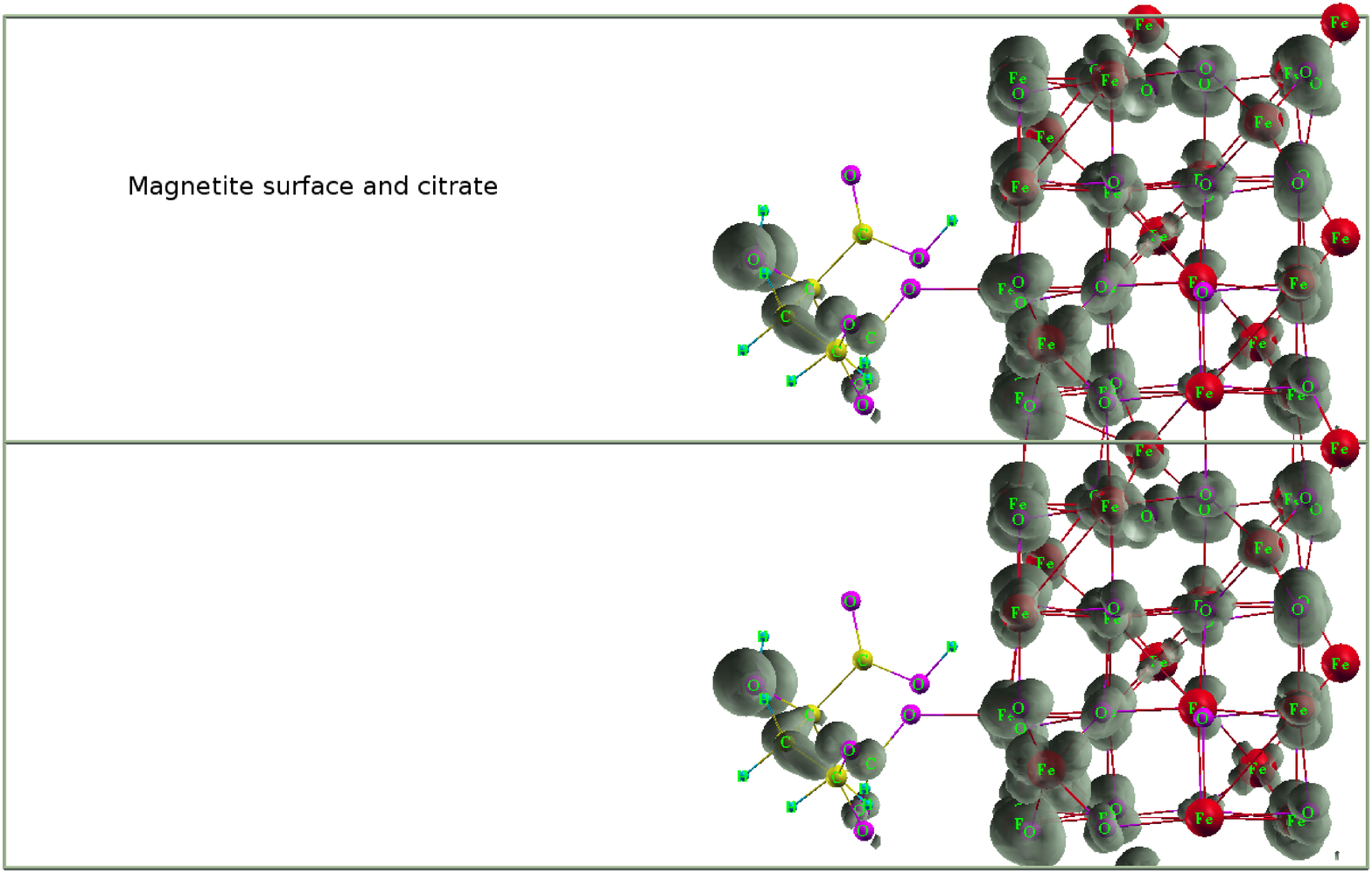}}
\caption{ \label{3dview} Three dimensional view the surfaces studied.
 Bonds are drawn using the default parameters of the XCrysden program, 
 and an electronic isosurface
  at the Fermi energy is drawn at 7\% of the maximum value in each case.}
							  
\end{figure}

We used standard ultrasoft pseudopotentials from the Quantum Espresso distribution using the
Perdew-Wang 91 functional in LDA+U~\cite{pw91}. Using this method, we could check 
that the magnetite crystal cell from~\cite{ocd,refocd} was structurally stable, 
, that we could reproduce the density of states from~\cite{pinto}, and that the magnetite
surface, using 3.4 nm of vacuum in the [001] direction, was subject to small structural changes
but that the corresponding total density of states was similar to the one of~\cite{Lodziana}. 
The magnetite surface is represented on figure~\ref{3dview} and the corresponding total density of states
on figure~\ref{totaldos}.

We first modeled the ligands we chose to study 
using the WebMO interface to the Gaussian09 code~\cite{gaussian09} 
using the Hartree-Fock method with the  
Hartree-Fock method with 6-31+G(d) basis set which is often considered as the best compromise between speed and 
accuracy in order to perform a quick structural optimization of such  molecules. 
We then checked that the obtained coordinates for the ligands corresponded to stable molecules
in the pseudopotential approach, added one of those molecules  at 1.5 nm of the magnetite surface
optimized as described above, and performed a full structural optimization going to 0 K by the standard
annealing method of the PWScf code. 

We chose two type of commonly used ligands : first, a $\pi$-electron donor type
for which we chose dopamine as commonly used in experimental work trying
to make nanoparticles biocompatible as described in the introduction, then
a $\sigma$-electron donor for which we chose citric acid, a molecule
also widely experimentally used~\cite{levy,Beigi}. A better description
would of course include water molecules or even charges in order to model
pH effects, but with about 500 active electrons and 100 atoms, as
well as 250000 G-vectors and no symmetry we feel that the system is at the
limit of what is nowadays computationally tractable.

\section{Results and discussion}

We present on figure~\ref{3dview} a view of three typical results
for the atomic positions.
On these results, it is clear that dopamine has a preferential 
adsorption site at the octahedral (A) iron atom of the magnetite
surface, when the citrate ligand has a preferential binding
on the tetrahedral (B) site of the magnetite surface. 

\begin{figure}
\includegraphics[scale=0.30,angle=0]{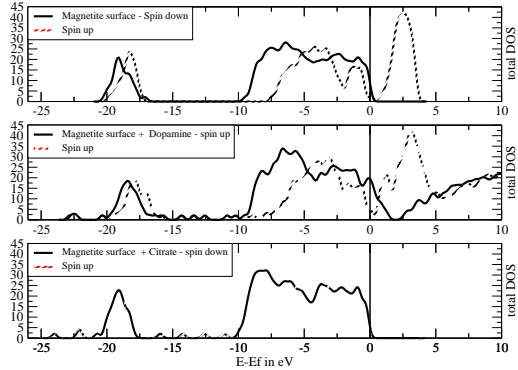} 
\caption{ \label{totaldos} Total density of states for the three chosen systems }
\end{figure}

This can be attributed to the presence of an aromatic cycle in dopamine
and $\pi$ electrons close to the hydroxyl group. From the total
density of states plotted on figure~\ref{totaldos}, one can see
that the presence of dopamine does change the small gap of magnetite
by adding some conduction electrons, when the presence of citrate
does not significantly changes the total density of states.
The difference of spin up and down density of states 
led us to suspect an effect of functionalization on magnetization
of the systems.

Those results are summarized on table~\ref{tbresults}.
Functionalization leads to a marked increase in magnetism, when
the value of magnetite surface alone is close to the one obtained 
by~\cite{Lodziana}. This increase in magnetic momenta can
be compared to results recently experimentally obtained by~\cite{li}.

Besides, in order to elucidate the role of $d$ orbitals on iron atoms on those
effects, we plot on figure~\ref{3dview} typical electronic densities isosurfaces at Fermi energy. 
One can see on those figures the typical $\pi$ character of orbitals contributed
by dopamine to conduction electrons at the surface, versus the lack of 
contribution of the citrate ligand. In both cases, there is however
a change of the $d$ character of the electrons contributing to conductivity
close to the surface, when in the case of the non-functionalized
 \begin{figure}
\includegraphics[scale=0.3,angle=0]{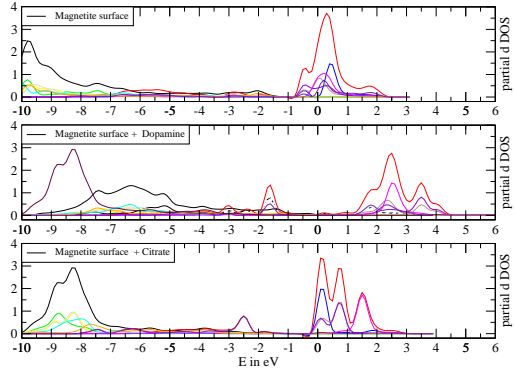} 
\caption{ \label{partialdosatom1} Partial density of states projected on a "d" state for an atom of type (A) at the magnetite
 surface, where dopamine preferentially binds}
\end{figure}
magnetite surface the conductivity rather comes from bulk electrons.
Those results remind us  of  those obtained by~\cite{Parkinson} about the change
in the conducting behavior of magnetite induced by hydrogen adsorption,
turning from a semiconductor to a half-metal.

 \begin{figure}
\includegraphics[scale=0.3,angle=0]{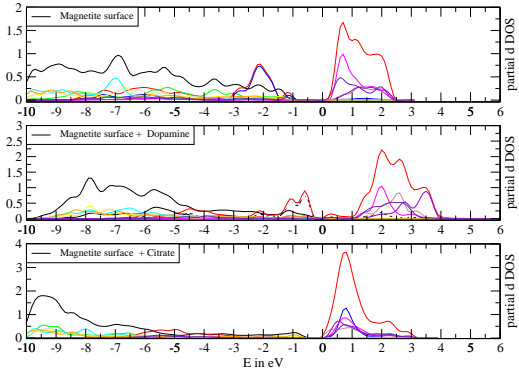} 
\caption{ \label{partialdosatom2} Partial density of states projected on a "d" state for an atom of type (B) at surface, where citrate
preferentially binds}
\end{figure}

In order to further analyze those results, we plot on figure~\ref{partialdosatom1}
and ~\ref{partialdosatom1}
projected densities of states for two typical atoms, namely Fe atoms 
of octahedral (A) types and tetrahedral (B) types where, respectively,
dopamine and citrate prefer to bind at the magnetite surface. Only
$d$-character wavefunctions are plotted around the chosen atoms,  
and one does see  that the presence of either dopamine or citrate
leads to a marked change in the projected densities of states around the chosen
atoms, the presence of dopamine shifting the positions of the peaks 
when the presence of citrate mainly changes the shapes of the peaks.
From this we conclude that the functionalization by dopamine will
induce a stronger change in the magnetic properties of the system
than the one by citrate, which, however, tends to induce a stronger~magnetization.

\begin{table}
    \begin{tabular}{|c|c|c|c|}
    \hline
    System                       & Magnetite s. & W.Citrate & W.Dopamine \\ \hline   
Fermi energy               & -1.2550 eV  & -0.4523 eV  & -0.6216 eV \\ \hline
Total mag.        & 75.03 $\mu_B$ /cell  & 92.69 $\mu_B$ /cell & 83.46 $\mu_B$ /cell \\\hline 
Absolute mag.       & 83.67 $\mu_B$ /cell  & 97.17 $\mu_B$ /cell & 90.90 $\mu_B$ /cell \\ 
    \hline
     \end{tabular}
\caption{Some quantitative results obtained on the chosen systems  \label{tbresults}}
\end{table}

\section{Conclusion}

In this paper, we address the effect of functionalization on the electronic
and magnetic properties of magnetite surface as an indicator of the
same properties in nanoparticles too big for a direct ab-initio approach.
Using well-established methods and references (namely LDA+U on
magnetite surfaces) we could verify that we had similar results to
 those
in recent papers on magnetite surfaces, 
and using two typical ligands, $\pi$ and $\sigma$ electron donors,
we could predict that those ligands would induce a different
change in the electronic properties of the systems, but in both cases 
an enhancement of magnetization. These findings are confirmed
by some recent experimental work. The present study could however
be improved by looking at temperature and solvent effects, by 
using a larger number of atoms in order to be closer to experimental
systems, as well as by modeling some other changes in physical properties
of magnetic nanoparticles of medical interest induced by surface functionalization.

\acknowledgments{In the course of this work, we used the CRIHAN
computing center as well as GENCI grant x2011096171
We thank J.-M.Gren\`eche and S.Ammar for fruitful discussions.
}

\end{document}